\documentclass[11pt]{article}
\usepackage{graphicx}
\usepackage{amssymb}
\usepackage{amsmath}
\usepackage[left]{lineno}

\begin{document}
\centerline{\LARGE EUROPEAN ORGANIZATION FOR NUCLEAR RESEARCH}
\vspace{15mm}
{\flushright{
\today \\
 }}
\vspace{15mm}

 \begin{center}

 {\bf {\Large \boldmath{Search for Heavy Neutrinos in $K^+ \rightarrow \mu^+ \nu_{\mu}$ Decays}}}
 \end{center}
\unboldmath
 
 \begin{center}
{\Large The NA62 Collaboration$\,$\renewcommand{\thefootnote}{\fnsymbol{footnote}}%
\footnotemark[1]\renewcommand{\thefootnote}{\arabic{footnote}}}\\
\end{center}
 
The NA62 experiment recorded a large sample of $K^+ \rightarrow \mu^+ \nu_{\mu}$ decays in 2007.
A peak search has been performed in the reconstructed missing mass
spectrum. 
In the absence of a signal,
limits in the range $2 \times 10^{-6}$ to $10^{-5}$ have been set on the squared mixing matrix element  $|U_{\mu4} |^2$ between muon and heavy neutrino states, 
for heavy neutrino masses in the range \mbox{300--375~MeV/$c^2$}. The result extends the range of masses for which upper limits have been set on the value of
 $|U_{\mu4} |^2$ in previous production search experiments.
 
\noindent
\begin{center}
\it{Accepted for publication in (Physics Letters B)}
\end{center}

\newpage 
\begin{center}
{\Large The NA62 collaboration}\
\vspace{2mm}

 C.~Lazzeroni$\renewcommand{\thefootnote}{\fnsymbol{footnote}}\footnotemark[1]\renewcommand{\thefootnote}{\arabic{footnote}}^,\,$\footnotemark[1],
 N.~Lurkin$\renewcommand{\thefootnote}{\fnsymbol{footnote}}\footnotemark[1]\renewcommand{\thefootnote}{\arabic{footnote}}^,\,$\footnotemark[2],
 F.~Newson,
 A.~Romano\\
{\em \small University of Birmingham, Edgbaston, Birmingham, B15 2TT, United Kingdom} \\[0.2cm]
 A.~Ceccucci,
 H.~Danielsson,
 V.~Falaleev,
 L.~Gatignon,
 S.~Goy Lopez$\,$\footnotemark[3], \\
 B.~Hallgren$\,$\footnotemark[4],
 A.~Maier,
 A.~Peters,
 M.~Piccini$\,$\footnotemark[5],
 P.~Riedler\\
{\em \small CERN, CH-1211 Gen\`eve 23, Switzerland} \\[0.2cm]
 P.L.~Frabetti,
 E.~Gersabeck$\,$\footnotemark[6],
 V.~Kekelidze,
 D.~Madigozhin,
 M.~Misheva$\,$\footnotemark[7], \\
 N.~Molokanova,
 S.~Movchan,
 Yu.~Potrebenikov,
 S.~Shkarovskiy,
 A.~Zinchenko$\,$\renewcommand{\thefootnote}{\fnsymbol{footnote}}\footnotemark[2]\renewcommand{\thefootnote}{\arabic{footnote}}\\
{\em \small Joint Institute for Nuclear Research, 141980 Dubna (MO), Russia} \\[0.2cm]
 P.~Rubin$\,$\footnotemark[8]\\
{\em \small George Mason University, Fairfax, VA 22030, USA} \\[0.2cm]
 W.~Baldini,
 A.~Cotta Ramusino,
 P.~Dalpiaz,
 M.~Fiorini,
 A.~Gianoli, \\
 A.~Norton,
 F.~Petrucci,
 M.~Savri\'e,
 H.~Wahl\\
{\em \small Dipartimento di Fisica e Scienze della Terra dell'Universit\`a e Sezione dell'INFN di Ferrara, \\I-44122 Ferrara, Italy} \\[0.2cm]
 A.~Bizzeti$\,$\footnotemark[9],
 F.~Bucci,
 M.~Veltri$\,$\footnotemark[10]\\
{\em \small Sezione dell'INFN di Firenze, I-50019 Sesto Fiorentino, Italy} \\[0.2cm]
 E.~Iacopini,
 M.~Lenti\\
{\em \small Dipartimento di Fisica dell'Universit\`a e Sezione dell'INFN di Firenze, I-50019 Sesto Fiorentino, Italy} \\[0.2cm]
 A.~Antonelli,
 M.~Moulson,
 M.~Raggi$\,$\footnotemark[11],
 T.~Spadaro\\
{\em \small Laboratori Nazionali di Frascati, I-00044 Frascati, Italy} \\[0.2cm]
 K.~Eppard,
 M.~Hita-Hochgesand,
 K.~Kleinknecht,
 B.~Renk,
 R.~Wanke,
 A.~Winhart$\,$\footnotemark[4]\\
{\em \small Institut f\"ur Physik, Universit\"at Mainz, D-55099 Mainz, Germany$\,$\footnotemark[12]} \\[0.2cm]
 R.~Winston\\
{\em \small University of California, Merced, CA 95344, USA} \\[0.2cm]
 V.~Bolotov$\,$\renewcommand{\thefootnote}{\fnsymbol{footnote}}\footnotemark[2]\renewcommand{\thefootnote}{\arabic{footnote}},
 V.~Duk$\,$\footnotemark[5],
 E.~Gushchin\\
{\em \small Institute for Nuclear Research, 117312 Moscow, Russia} \\[0.2cm]
 F.~Ambrosino,
 D.~Di Filippo,
 P.~Massarotti,
 M.~Napolitano,
 V.~Palladino$\,$\footnotemark[13],
 G.~Saracino\\
{\em \small Dipartimento di Fisica dell'Universit\`a e Sezione dell'INFN di Napoli, I-80126 Napoli, Italy} \\[0.2cm]
 G.~Anzivino,
 E.~Imbergamo,
 R.~Piandani$\,$\footnotemark[14],
 A.~Sergi$\,$\footnotemark[4]\\
{\em \small Dipartimento di Fisica dell'Universit\`a e Sezione dell'INFN di Perugia, I-06100 Perugia, Italy} \\[0.2cm]
 P.~Cenci,
 M.~Pepe\\
{\em \small Sezione dell'INFN di Perugia, I-06100 Perugia, Italy} \\[0.2cm]
 F.~Costantini,
 N.~Doble,
 S.~Giudici,
 G.~Pierazzini$\,$\renewcommand{\thefootnote}{\fnsymbol{footnote}}\footnotemark[2]\renewcommand{\thefootnote}{\arabic{footnote}},
 M.~Sozzi,
 S.~Venditti\\
{\em \small Dipartimento di Fisica dell'Universit\`a e Sezione dell'INFN di Pisa, I-56100 Pisa, Italy} \\[0.2cm]
 S.~Balev$\,$\renewcommand{\thefootnote}{\fnsymbol{footnote}}\footnotemark[2]\renewcommand{\thefootnote}{\arabic{footnote}},
 G.~Collazuol$\,$\footnotemark[15],
 L.~Di Lella,
 S.~Gallorini$\,$\footnotemark[15],
 E.~Goudzovski$\,$\footnotemark[1]$^,\,$\footnotemark[2]$^,\,$\footnotemark[4], \\
 G.~Lamanna$\,$\footnotemark[16],
 I.~Mannelli,
 G.~Ruggiero$\,$\footnotemark[17]\\
{\em \small Scuola Normale Superiore e Sezione dell'INFN di Pisa, I-56100 Pisa, Italy} \\[0.2cm]
 C.~Cerri,
 R.~Fantechi\\
{\em \small Sezione dell'INFN di Pisa, I-56100 Pisa, Italy} \\[0.2cm]
 S.~Kholodenko,
 V.~Kurshetsov,
 V.~Obraztsov,
 V.~Semenov,
 O.~Yushchenko\\
{\em \small Institute for High Energy Physics, 142281 Protvino (MO), Russia$\,$\footnotemark[18]} \\[0.2cm]
 G.~D'Agostini\\
{\em \small Dipartimento di Fisica, Universit\`a di Roma La Sapienza e \\Sezione dell'INFN di Roma I, I-00185 Roma, Italy} \\[0.2cm]
 E.~Leonardi,
 M.~Serra,
 P.~Valente\\
{\em \small Sezione dell'INFN di Roma I, I-00185 Roma, Italy} \\[0.2cm]
 A.~Fucci,
 A.~Salamon\\
{\em \small Sezione dell'INFN di Roma Tor Vergata, I-00133 Roma, Italy} \\[0.2cm]
 B.~Bloch-Devaux$\,$\footnotemark[19],
 B.~Peyaud\\
{\em \small DSM/IRFU -- CEA Saclay, F-91191 Gif-sur-Yvette, France} \\[0.2cm]
 J.~Engelfried\\
{\em \small Instituto de F\'isica, Universidad Aut\'onoma de San Luis Potos\'i, 78240 San Luis Potos\'i, Mexico$\,$\footnotemark[20]} \\[0.2cm]
 D.~Coward\\
{\em \small SLAC National Accelerator Laboratory, Stanford University, Menlo Park, CA 94025, USA} \\[0.2cm]
 V.~Kozhuharov$\,$\footnotemark[21],
 L.~Litov\\
{\em \small Faculty of Physics, University of Sofia, BG-1164 Sofia, Bulgaria$\,$\footnotemark[22]} \\[0.2cm]
 R.~Arcidiacono$\,$\footnotemark[23],
 S.~Bifani$\,$\footnotemark[4]\\
{\em \small Dipartimento di Fisica dell'Universit\`a e Sezione dell'INFN di Torino, I-10125 Torino, Italy} \\[0.2cm]
 C.~Biino,
 G.~Dellacasa,
 F.~Marchetto\\
{\em \small Sezione dell'INFN di Torino, I-10125 Torino, Italy} \\[0.2cm]
 T.~Numao,
 F.~Reti\`ere\\
{\em \small TRIUMF, Vancouver, British Columbia, V6T 2A3, Canada} \\[0.2cm]

\end{center}
%
%
\renewcommand{\thefootnote}{\fnsymbol{footnote}}
\footnotetext[1]{Corresponding author, email: cristina.lazzeroni@cern.ch, nicolas.lurkin@cern.ch}   
\footnotetext[2]{Deceased}
\renewcommand{\thefootnote}{\arabic{footnote}}

\footnotetext[1]{Supported by a Royal Society University Research Fellowship}
\footnotetext[2]{Supported by ERC Starting Grant 336581}
\footnotetext[3]{Present address: CIEMAT, E-28040 Madrid, Spain}
\footnotetext[4]{Present address: School of Physics and Astronomy, University of Birmingham, Birmingham, B15 2TT, UK}
\footnotetext[5]{Present address: Sezione dell'INFN di Perugia, I-06100 Perugia, Italy}
\footnotetext[6]{Present address: Ruprecht-Karls-Universit\"at Heidelberg, D-69120 Heidelberg, Germany}
\footnotetext[7]{Present address: Institute of Nuclear Research and Nuclear Energy of Bulgarian Academy of Science (INRNE-BAS), BG-1784 Sofia, Bulgaria}
\footnotetext[8]{Funded by the National Science Foundation under award No. 0338597}
\footnotetext[9]{Also at Dipartimento di Fisica, Universit\`a di Modena e Reggio Emilia, I-41125 Modena, Italy}
\footnotetext[10]{Also at Istituto di Fisica, Universit\`a di Urbino, I-61029 Urbino, Italy}
\footnotetext[11]{Present address: Universit\`a di Roma La Sapienza, I-00185 Roma, Italy}
\footnotetext[12]{Funded by the German Federal Minister for Education and Research (BMBF) under contract 05HA6UMA}
\footnotetext[13]{Present address: Physics Department, Imperial College London, London, SW7 2BW, UK}
\footnotetext[14]{Present address: Sezione dell'INFN di Pisa, I-56100 Pisa, Italy}
\footnotetext[15]{Present address: Dipartimento di Fisica dell'Universit\`a e Sezione dell'INFN di Padova, I-35131 Padova, Italy}
\footnotetext[16]{Present address: Dipartimento di Fisica dell'Universit\`a e Sezione dell'INFN di Pisa, I-56100 Pisa, Italy}
\footnotetext[17]{Present address: Department of Physics, University of Liverpool, Liverpool, L69 7ZE, UK}
\footnotetext[18]{Partly funded by the Russian Foundation for Basic Research grant 12-02-91513}
\footnotetext[19]{Present address: Dipartimento di Fisica dell'Universit\`a, I-10125 Torino, Italy}
\footnotetext[20]{Funded by Consejo Nacional de Ciencia y Tecnolog\'ia (CONACyT) and Fondo de Apoyo a la Investigaci\'on (UASLP)}
\footnotetext[21]{Also at Laboratori Nazionali di Frascati, I-00044 Frascati, Italy}
\footnotetext[22]{Funded by the Bulgarian National Science Fund under contract DID02-22}
\footnotetext[23]{Also at Universit\`a degli Studi del Piemonte Orientale, I-13100 Vercelli, Italy}

\clearpage
 
\section{Introduction}\label{sec:intro}

The fact that neutrinos oscillate implies that they have non-zero masses. 
While in the Standard Model (SM) neutrinos are massless by construction, the SM can be extended in various ways to accommodate neutrino masses~\cite{pal}. 
In a large class of models, the see-saw mechanism is used to explain the lightness of the SM neutrinos by introducing additional heavy neutrino mass states which mix with the SM flavour states~\cite{moh}. One example of models including heavy neutrinos is the neutrino minimal Standard Model ($\nu$MSM), in which three right-handed neutrinos are added to the SM with one of them being at the GeV scale~\cite{asaka, shap}.
For heavy neutrinos with masses 
below the kaon mass, 
limits on their mixing matrix elements can be placed 
by searching for peaks in the missing mass spectrum of  $K^\pm$ decays~\cite{shrock}.
In the following, two-body kaon decays to a muon and a SM neutrino are denoted $K^+ \rightarrow \mu^+ \nu_{\mu}$,
while those with a muon and a heavy neutrino are denoted $K^+ \rightarrow \mu^+ \nu_{h}$; the notation $K^+ \rightarrow \mu^+ N$ indicates either case.
Limits on $|U_{\mu 4}|^2$ in the extended neutrino mixing matrix 
using the process $K^+ \rightarrow \mu^+ \nu_h$
come from experiments with stopped kaons, and are of the order of $10^{-8}$ up to 300 MeV/$c^2$~\cite{hay} and $10^{-6}$ up to 330 MeV/$c^2$~\cite{arta}.

The ratio of the $K^+$ decay width to heavy neutrino
to the decay width to SM muon neutrinos is related to $|U_{\mu 4}|^2$~\cite{shrock}:

\begin{equation}
\frac{ \mathcal{B} (K^+ \rightarrow \mu^+ \nu_h ) }{ \mathcal{B} ( K^+ \rightarrow \mu^+ \nu_{\mu})} = |U_{\mu 4}|^2 
f (m_h) \, ,
\label{eq:formula}
\end{equation}

\noindent where $m_h$ is the mass of the heavy neutrino, and $f (m_h)$ accounts for the phase space factor
and the helicity suppression, and varies in the range 1.5--4.0 for $m_h$ in the region 300--375 MeV/$c^2$ considered in the present analysis.

Under the assumption that heavy neutrinos decay only to SM particles, the lifetime of a heavy neutrino is determined by the mixing matrix elements 
and by its mass~\cite{gorb}. 
For heavy neutrino masses in the range 300--375 MeV/$c^2$, the dominant decay modes are $\nu_h \rightarrow \pi^0 \nu_{e, \mu, \tau}$ and 
$\nu_h \rightarrow \pi^+ \ell^-$, where $\ell  = e, \mu$.
Assuming $|U_{\ell 4}|^2 < 10^{-4}$ with $\ell= e, \mu, \tau$,
the mean free path of heavy neutrinos at NA62 for any mass in the range considered is greater than 10 km, and therefore
their decays can be neglected, since the probability of decaying in the detector or decay volume is below 1\%.

\section{Beam, detector and data samples}

The beam line and detector of the earlier NA48/2 experiment were reused by the NA62 experiment during $2007$ data taking;
they are described in detail in~\cite{detector, beam}. 
Primary protons of 400 GeV/$c$, extracted from the CERN SPS, impinged on a 40 cm long, 0.2 cm diameter 
beryllium target. 
Secondary beams of positively and negatively charged hadrons were produced, momentum-selected,
similarly focused and transported to the detector. These beams could be run simultaneously or separately.
The central beam momentum of 74 GeV/$c$ was selected by the first two
magnets in a four-dipole achromat and by momentum-defining slits incorporated into a 3.2 m
thick copper/iron proton beam dump, which also provided the
possibility of blocking either of the two beams. 
The beams had a momentum spread of $\pm 1.4$~GeV/$c$ (rms).
For about $1.8 \times 10^{12}$ primary protons incident on the target per SPS spill of 4.8~s duration, 
the secondary beam fluxes at the entrance to the decay volume were, respectively, $1.7 \times 10^7$ and 
$0.8 \times 10^7$ positively and negatively charged particles per spill.

The fraction of kaons in each beam was about 6\%. The beam kaons decayed in a fiducial volume contained in a 114 m long cylindrical evacuated tank. 
The $K^+$ and $K^-$ beams were deflected at the entrance of the fiducial volume by angles in the range $\pm (0.23-0.30)$ mrad with respect to
the longitudinal $z$ axis to compensate for the opposite $\mp 3.58$ mrad deflections by the
spectrometer magnet. All these deflections were regularly reversed during the data taking.
The hadron beams were accompanied by a flux of stray (halo) muons produced by kaon and pion decays upstream of the decay volume.
Two 
magnetized iron toroids with apertures centred on the beam line were installed upstream of the decay volume to deflect positive halo  muons and suppress the associated backgrounds.

Charged particle momenta were measured by a magnetic spectrometer, housed in a tank filled with helium at approximately atmospheric pressure, placed downstream of the decay volume. The spectrometer consisted of four drift chambers (DCHs), 
each comprising four views of double planes of staggered sense wires, and a dipole magnet located between the second and the third DCHs. The magnet provided a horizontal transverse momentum kick of 265~MeV/$c$ to charged particles, and the spectrometer had a momentum resolution of
$\sigma_p /p = 0.48\% \oplus 0.009\% \cdot p$, where the momentum $p$ is expressed in GeV/$c$. 
A hodoscope (HOD) consisting of two planes of plastic scintillator strips producing fast trigger signals was placed after the spectrometer.
A liquid krypton (LKr) electromagnetic calorimeter of thickness 127 cm (27$X_0$) was located further downstream. Its 13248 readout cells had a transverse size of $2\times2$ cm$^2$ each with no longitudinal segmentation. The energy resolution was 
$\sigma_E /E = 3.2\% / \sqrt{E} \oplus 9\% / E \oplus 0.42\%$, and the spatial resolution for the transverse coordinates $x$ and $y$ of an isolated electromagnetic shower was $\sigma_x = \sigma_y = 0.42$ cm $/ \sqrt{E} \oplus 0.06 $ cm, where $E$ is expressed in GeV.
A muon detector (MUV) was located further downstream. The MUV was composed of three planes of plastic scintillator strips (aligned horizontally in the first and last planes, and vertically in the middle plane) read out by photomultipliers at both ends. Each strip was $2.7$ m long and 1 cm thick. The widths of the strips were 25 cm in the first two planes, and $45$ cm in the third plane. The MUV was preceded by a hadronic calorimeter (6.7 nuclear interaction lengths) not used for the present measurement.
Each MUV plane was preceded by an additional 0.8 m thick iron absorber.

General data taking conditions are described in~\cite{rk}.
The main trigger condition for selecting the sample of $K^+ \rightarrow \mu^+ N$ decays required the coincidence in time and space of signals in the two HOD planes (HOD signal), and loose lower and upper limits on the DCH hit multiplicity 
(1-track signal), downscaled by a factor of 150.  
Data taking periods with simultaneous beams were collected with a lead bar installed between the two HOD planes 
for muon identification studies.
For data collected with the lead bar in place, the
vetoing power for backgrounds with photons is reduced, so these data are excluded from
the present analysis. 
Since the muon halo background is smaller in the $K^+$ sample, this analysis is based on data with the $K^+$ beam only
(43\% of the integrated kaon flux, as used in~\cite{rk1}),
while data taken with only the $K^-$ beam are used to study the background from halo muons.

\section{Analysis strategy}\label{strategy}

In the decay $K^+ \rightarrow \mu^+ N$ the neutrino mass can be reconstructed 
as $m^2_{h} = m^2_{\text{miss}} = (p_K - p_{\mu} )^2$, where
$p_K$ and $p_{\mu}$ are the four-momenta of the kaon and the muon respectively.
The kaon momentum is not measured on an event-by-event basis, and $p_K$ is obtained, assuming the kaon mass, from the average three-momentum measured with $K^+ \rightarrow \pi^+ \pi^+ \pi^-$ decays approximately every 500 SPS spills.
The muon four-momentum $p_{\mu}$ is determined as that of a reconstructed charged track, assumed to be a muon. 

Simulated samples of $K^+ \rightarrow \mu^+ \nu_h$ decays, with mass $m_h$ between 240 and 380 MeV/$c^2$ 
at 1 MeV/$c^2$ intervals, have been generated to determine the signal acceptance.
The kaon decay modes that contribute to the background have been simulated to determine the expected spectrum of the reconstructed $m_{\text{miss}}$. The contribution to the background from muon
halo is evaluated using a control data sample, defined as the sample recorded with the $K^-$ beam only.
The integrated $\mu^+$ halo flux in the control sample is three times smaller than that 
in the $K^+$ sample.

 A comparison between the expected and observed $m_{\text{miss}}$ spectra is used to set limits on the observed number of $K^+ \rightarrow \mu^+ \nu_h$ decays  for each assumed $\nu_{h}$ mass.  These are translated into limits on the branching ratio $\mathcal{B} (K^+ \rightarrow \mu^+ \nu_h) $ and the mixing parameter $|U_{\mu 4}|^2$.
The search for heavy neutrinos is restricted to the mass range 300--375 MeV/$c^2$ (called the signal region in the following).
Lower neutrino masses are accessible, but a strong limit on $|U_{\mu4}|^2$ of the order of $10^{-8}$ from a production experiment exists below 300 MeV/$c^2$~\cite{arta}.
The reconstructed missing mass range 245--298 MeV/$c^2$, where the $\nu_h$ presence is excluded by this limit, is used as a control region to measure the trigger efficiency for the background events.   

\section{Event selection}\label{selection}

Charged particle trajectories and momenta are reconstructed from hits and drift times in the spectrometer using a detailed magnetic field map.
The reconstructed $K^+ \rightarrow \pi^+ \pi^+ \pi^-$ invariant mass is used for fine calibration of the spectrometer momentum scale and DCH alignment throughout the data taking. 
Clusters of energy deposition in the LKr calorimeter are found by locating maxima in space and time in the digitized pulses from individual cells. 
Reconstructed energies are corrected for energy outside the cluster boundaries, energy lost in isolated inactive cells (0.8\% of the total number), sharing of energy between clusters, and non-linearity for clusters with energy below 11 GeV.

The selection requires exactly one positively charged track with the following characteristics:
within the DCH, LKr calorimeter and MUV geometrical acceptance; momentum $p$ between 10 and 65 GeV/$c$;
within 20 ns of  the trigger time recorded by the HOD; 
distance of closest approach (CDA)
between the track and the beam axis, as monitored with $K^+ \rightarrow \pi^+ \pi^+ \pi^-$ decays, smaller than 3~cm; 
track extrapolation 
associated in time and space with MUV signals from the first two planes.

Selected events are required to be free of clusters of energy deposition in the LKr calorimeter except 
for any of the following configurations: the cluster energy is lower than 2~GeV; the cluster time is more 
than 12~ns away from the track time; the cluster is consistent with bremsstrahlung from the track before deflection by the spectrometer magnet (within 6~cm of the straight-line extrapolated upstream track); the cluster position is within 40~cm of the extrapolated downstream track.

\section{Background contributions}\label{background}

The background receives contributions from muon halo, evaluated with the control data sample,
and from kaon decays, evaluated with simulation.

\subsection*{Muon halo background}\label{halo}

A data driven approach is used in modelling the muon halo contribution, and
in designing a selection that minimizes this background while preserving signal acceptance. 
The distribution of halo background events is estimated using the control sample (see Section~\ref{strategy}).
The majority of reconstructed $\mu^+$ in the control sample comes from muon halo with two sources of contamination: 1)  $K^+$ in specific momentum bands pass through the beam absorbers (with a probability of up to $5 \times 10^{-4}$ depending on momentum) and decay into $K^+ \rightarrow \mu^+ \nu_{\mu}$; 
a simulation shows that the reconstructed $m_{\text{miss}}$ calculated
assuming the nominal kaon momentum is lower than 280 MeV/$c^2$, and therefore this component
does not enter the signal region; 2) the contribution from mis-identified positively charged pions from $K^- \rightarrow \pi^- \pi^- \pi^+$ decays enters the signal region,
and is simulated and subtracted.

\begin{figure}[htb]
\begin{minipage}[t]{18pc}
\includegraphics[width=18pc]{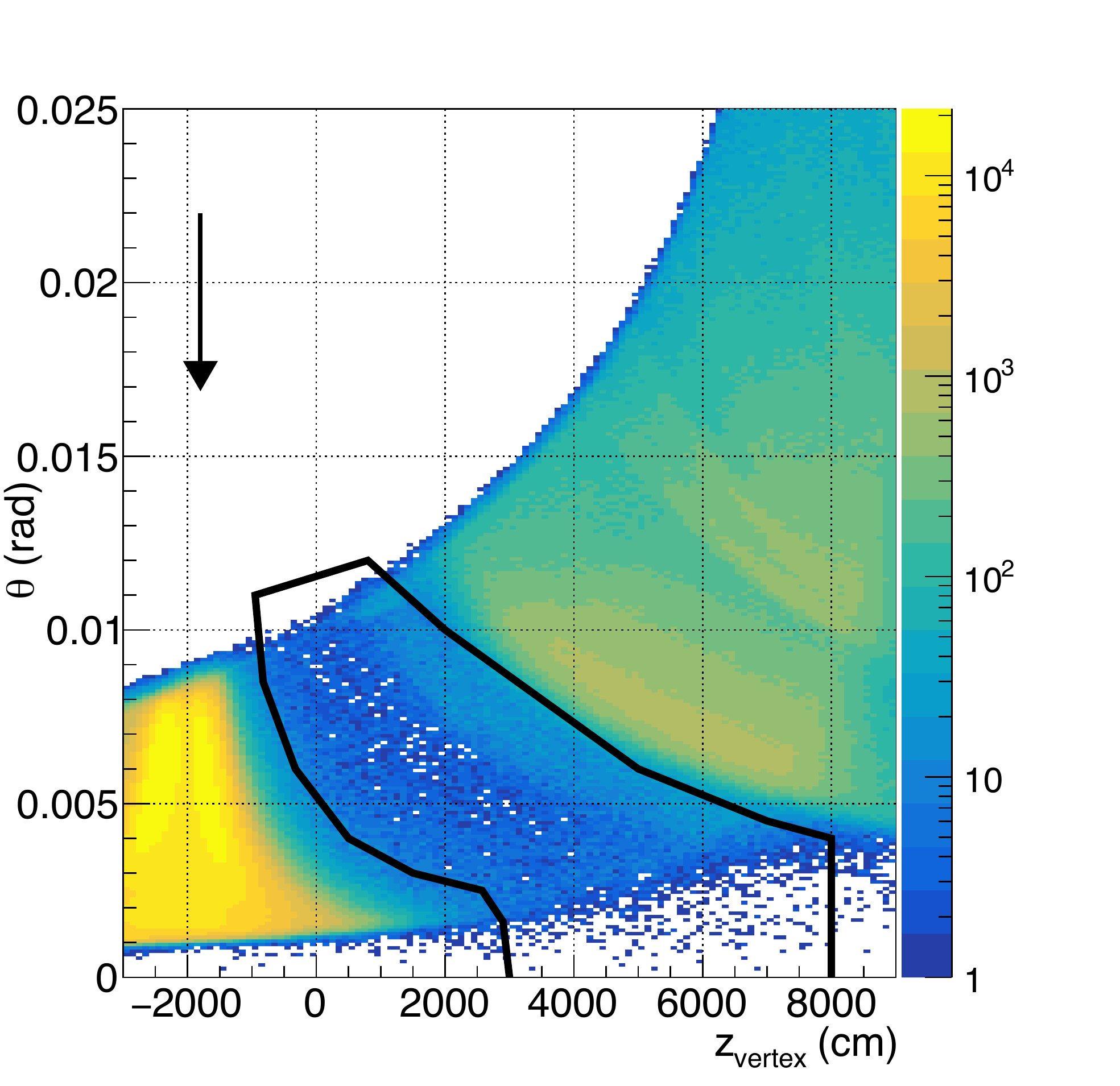}
\end{minipage}\hspace{2pc}%
\begin{minipage}[t]{18pc}
\includegraphics[width=18pc]{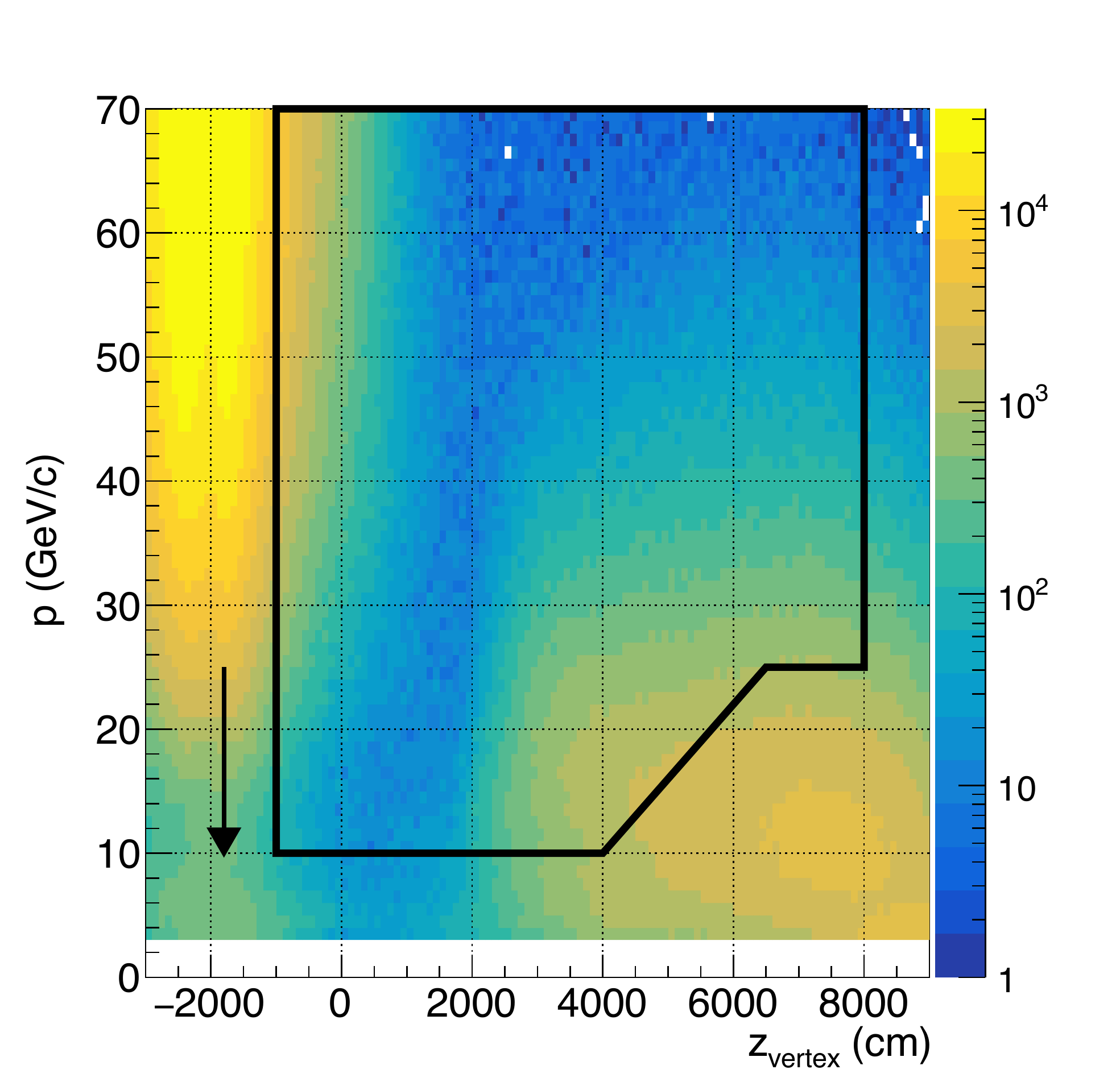}
\end{minipage}
\caption{
Distribution of halo events in the ($z_{\text{vertex}},\theta$) plane (left) and ($z_{\text{vertex}}, p$) plane (right). The contours show the projections of the five-dimensional selection criteria. The events outside the contours are rejected. The arrow indicates the start of the fiducial volume. See Section~\ref{background} for details.}
\label{fig:halo_regions}
\end{figure}

To study the halo, the event selection described in Section~\ref{selection} is used with a relaxed CDA condition (CDA $<8$ cm).
The distribution of the events in the control sample passing this selection is shown in Figure~\ref{fig:halo_regions} in the 
variables $z_{\text {vertex}}$, track momentum $p$ and $\theta$, where $\theta$ is the angle between the $K^+$ beam axis and the measured muon direction. 
To mimimize the halo contribution, additional selection criteria are applied in a five-dimensional space ($z_{\text{vertex}}, \theta, p, \text{CDA}, \phi$), where
$\phi$ is the track azimuthal angle in the transverse plane. The contours in  Figure~\ref{fig:halo_regions} show example 
projections of these five-dimensional criteria;
the events outside the contours are rejected. 
The signal acceptance reduction due to the multi-dimensional criteria with respect to the selection described in Section~\ref{selection} is in the range 40--45\% depending on $m_h$.

The estimated number of halo background events in the final sample is obtained from the number of events observed in the control sample, 
normalized to the $K^+$ data in the range $m^2_{\text{miss}} > 0.05$~GeV$^2$/$c^4$
and $3<$ CDA $< 8$~cm. 

\subsection*{Kaon decay background}

\begin{figure}[htb]
\begin{minipage}[t]{18pc}
\includegraphics[width=18pc]{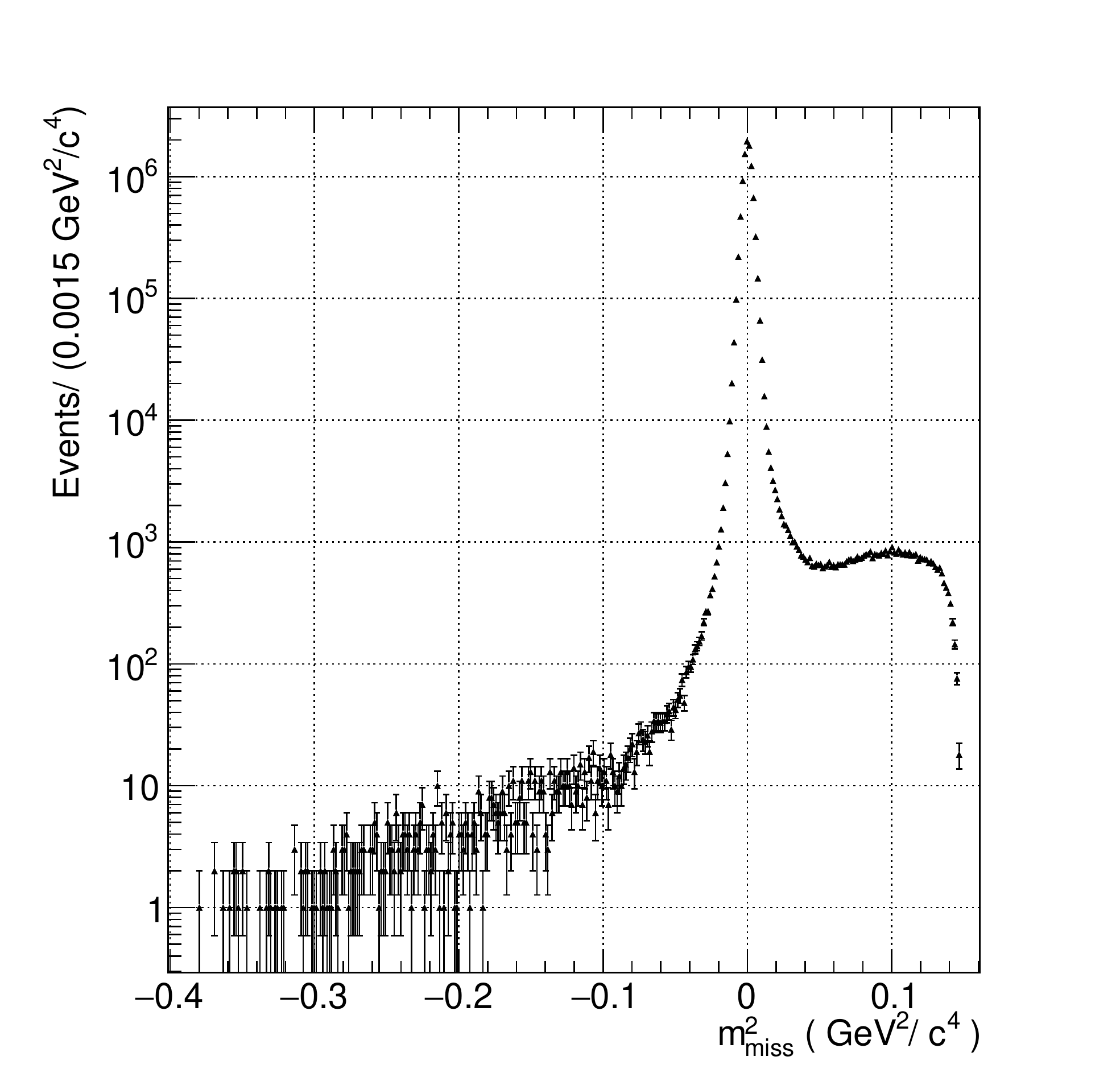}
\caption{ Reconstructed squared missing mass distribution for data passing the final event selection.}
\label{full_mass}
\end{minipage}\hspace{2pc}%
\begin{minipage}[t]{18pc}
\includegraphics[width=18pc]{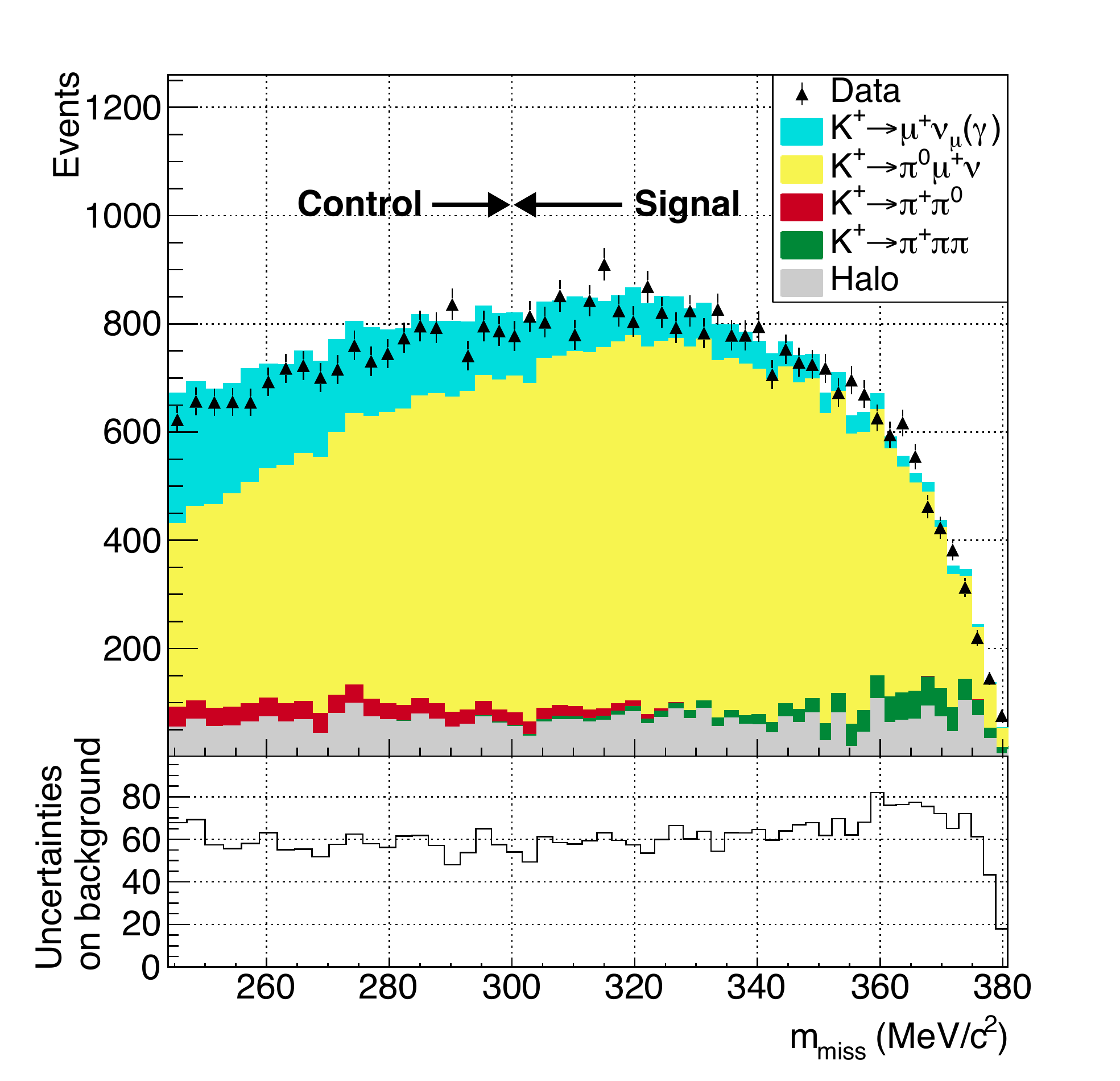}
\caption{\label{fig:bgcomp} Missing mass distributions for data, showing statistical uncertainties, and for the estimated background contributions, 
in both signal and control regions. 
The lower plot shows the total uncertainty on the background estimate.}
\end{minipage}
\end{figure}

The total number of kaon decays in the fiducial region, $N_K$, 
is used to scale the simulated distributions of the expected backgrounds. It is measured with a sample of $K^+ \rightarrow \mu^+ \nu_{\mu}$ decays using the selection described in \cite{rk} after adding the kinematic criteria;
the number of events in the missing mass squared distribution within $ |m^2_{\text{miss}} | < 0.015$ GeV$^2 /c^4$ is evaluated 
after subtracting a sub-percent contribution from beam halo.  The squared missing mass distribution is shown in Figure~\ref{full_mass}. 
The number of $K^+ \rightarrow \mu^+ \nu_{\mu}$ decays after background subtraction is $9.45 \times 10^6$ and the corresponding acceptance is 24.88\%.
The resulting number of kaon decays in the fiducial volume in the analysed dataset  
is $N_K =  (5.977 \pm 0.015) \times 10^7$.
 

The decay $K^+ \rightarrow \mu^+ \nu_{\mu} $ 
forms a peak at zero $m^2_{\text{miss}}$ with a width 
determined by the width of the kaon momentum spectrum,
the beam divergence and the spectrometer resolution; the peak is well outside the 300--375 MeV/$c^2$ signal region.
The contribution from $K^+ \rightarrow \mu^+ \nu_{\mu} \gamma$ decay appears as a high-mass tail in the $m^2_{\text{miss}}$ distribution and is taken into account by the simulation.
The dominant background from kaon decays in the signal region comes from $K^+ \rightarrow \pi^0 \mu^+ \nu_{\mu} $ decays with an undetected $\pi^0$ due to the non-hermetic geometrical acceptance.
The hadronic decay $K^+ \rightarrow \pi^+ \pi^0$ is only reconstructed as signal if the $\pi^0$
is undetected and the $\pi^+$ is mis-identified as a muon or decays into a muon.

The backgrounds due to kaon decays to three pions are naturally suppressed because they involve either three tracks ($K^+ \rightarrow \pi^+ \pi^+ \pi^-$) or photons  ($K^+ \rightarrow \pi^+ \pi^0 \pi^0$).
The events which pass the selection typically appear at the upper end of the $m^2_{\text{miss}}$ spectrum.
Decays with positrons in the final state ($K^+ \rightarrow e^+ \nu_e$,  $K^+ \rightarrow \pi^0 e^+ \nu_e$) are rejected with particle identification.

\section{Systematic uncertainties on the background estimate}\label{sec:sys}

The uncertainty on kaon decay background receives contributions from the uncertainty on the number of kaon decays in the fiducial volume, $N_K$, and the individual kaon decay branching ratios. The contribution from the finite size of the simulated samples is negligible. 
The estimate of the systematic uncertainty associated with $N_K$ is obtained by varying the cut on $|m^2_{\text{miss}}|$ by $\pm \sigma_{m^2}$ resulting in a variation of 0.2\% (where $ \sigma_{m^2}=3.1 \times 10^{-3}$ GeV$^2 /c^4$ is the resolution on the $K^+ \rightarrow \mu^+ \nu_{\mu}$ signal peak). The contribution from  $\mathcal{B}(K^+ \rightarrow \mu^+ \nu_{\mu})$ results in a variation of 0.15\%.
The overall systematic uncertainty on kaon decay background varies from 0.6\% to 1.0\% of the total expected background as a function of $m^2_{\text{miss}}$, and it is dominated in the signal region by the uncertainty on $\mathcal{B}(K^+ \rightarrow \pi^0 \mu^+ \nu)$.

To estimate the uncertainty on the $K^+ \rightarrow \mu^+ \nu_{\mu}$ background due to non-Gaussian tails in the DCH resolution, 
a sample of $K^+ \rightarrow \pi^+ \pi^0 \,\, (\pi^0 \rightarrow \gamma \gamma) $ decays, selected with LKr calorimeter without the spectrometer information, is used. 
The expected $\pi^+$ three-momentum is computed from the photon cluster energies and positions using the average kaon momentum
and $\pi^0$ mass constraint.
Data and simulated  $K^+\to \pi^+ \pi^0$ are then compared in terms of
the agreement between the measured and expected missing mass, $\pi^+$ momentum and direction spectra.
From this comparison it is inferred that the uncertainty on the background estimate in the $K^+ \rightarrow \mu^+ \nu_{h}$ signal region does not exceed 6\%
of the total expected background; this uncertainty affects mostly the low $\nu_h$ mass region.

The systematic uncertainty attributed to the halo background arises from the limited size of the control sample (halo statistical contribution), and 
from the assumption that the halo distribution in the control sample accurately reproduces that of the $K^+$ data (halo model contribution).
The halo statistical contribution is 2--4\% of the total expected background in the range 300--360 MeV/$c^2$ and rises to 16\% in the range 360--375 MeV/$c^2$.
The control sample is divided into sub-samples according to 
selection variables and each sub-sample is normalised to the $K^+$ data.
The halo model contribution is evaluated by comparing the normalizations obtained with the different sub-samples with that obtained for the entire sample.
This contribution is 1--3\% of the total expected background in the range 300--360 MeV/$c^2$ and rises to 8\% in the range 360--375 MeV/$c^2$.
The uncertainty due to the subtraction of  $K^- \rightarrow \pi^- \pi^- \pi^+$ events is negligible.

A $K^+ \rightarrow \mu^+ \nu_{\mu}$ sample is used to measure the MUV muon identification   
efficiency as a function of track momentum. 
This efficiency varies between 96\% and 98\% over the momentum range between 10 and 65 GeV/$c$.
The simulation is tuned to reproduce this efficiency to 1\% precision, and therefore a systematic uncertainty of 1\%
is assigned to the total expected background.

The HOD trigger inefficiency is $(1.4 \pm 0.1)\%$ as discussed in \cite{rk}; 
since the inefficiency depends mainly on the number of tracks which is the same for signal, $K^+ \rightarrow \mu^+ \nu_{\mu}$ decays and main backgrounds,
it cancels out to a good approximation.  
The 1-track trigger inefficiency for $K^+ \rightarrow \mu^+ \nu_{\mu}$ decays was measured with respect to the HOD trigger
to be much smaller that the HOD inefficiency~\cite{rk}
and can be neglected.
Conversions of undetected photons from $K^+ \rightarrow \pi^0 \mu^+ \nu$ decays cause a 1-track inefficiency due to events 
with high multiplicity of hits in the DCH chambers.
The 1-track trigger efficiency for the background could be evaluated directly in the signal region, or by extrapolating the 
measurement performed in the control region to the signal region.
However, the possible presence of  $K^+ \rightarrow \mu^+ \nu_{h}$ decays in the signal region would increase
the apparent efficiency, thereby affecting the signal sensitivity.
Therefore the 1-track trigger efficiency for the background is evaluated in the control $m_{\text{miss}}$ region 245--298~MeV/$c^2$, since strong limits on the heavy neutrino production in this region already exist.
In this control region the 1-track efficiency is $(89.8\pm0.6) $\% and was shown not to depend on the missing mass. 
The uncertainty on the trigger efficiency for the background translates into a contribution of 0.7\% on the total expected background.

\section{Upper limits on heavy neutrino production}\label{sec:limits}

The event selection described in Section~\ref{selection} with the addition of 
the five-dimensional criteria described in Section~\ref{halo}
constitutes the final selection.
Figure~\ref{fig:bgcomp} shows the $m_{\text{miss}}$ distribution of events passing the final selection and the estimated background spectrum.
The halo contribution varies as a function of $m_{\text{miss}}$ between 5\% and 20\% of the background and carries the largest relative systematic uncertainty.

\begin{figure}[htb]
\begin{center}
\includegraphics[width=18pc]{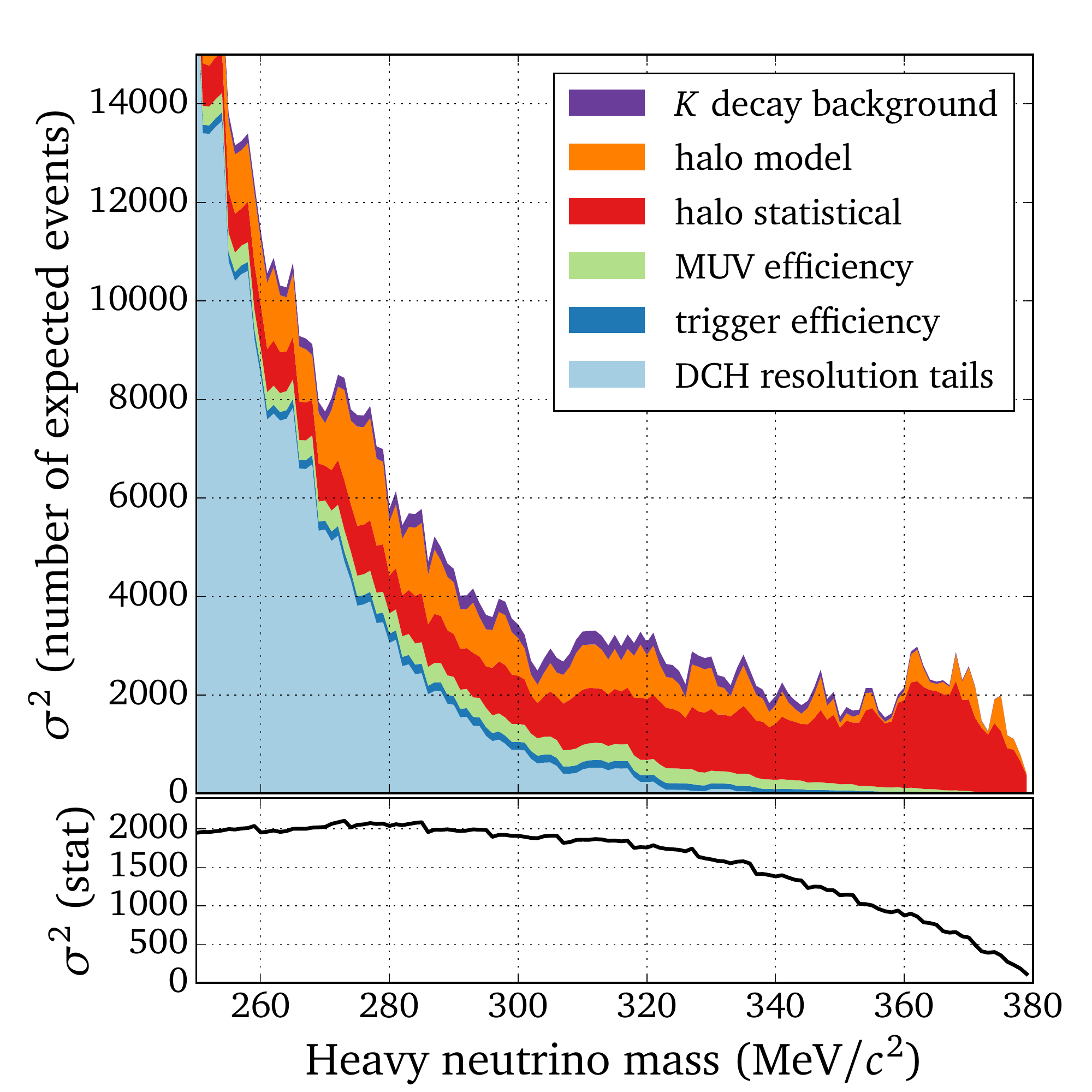}
\caption{Upper plot: squared uncertainties on the numbers of expected background events at each heavy neutrino mass. Lower plot: squared statistical uncertainty for data.}
\label{fig:syst}
\end{center}
\end{figure}

For each neutrino mass $m_h$ under consideration in the signal region 300--375 MeV/$c^2$, a window 
of $\pm \sigma_h$ in the missing mass spectrum is defined centred on $m_h$, 
where $\sigma_h$ is the resolution parametrized as $\sigma_h =$ 12 MeV/$c^2 - 0.03\cdot m_h$.
For each window, the width is rounded to the nearest multiple of $10^{-4}$ GeV$^2$/$c^4$ in $m^2_{\text{miss}}$.
The signal acceptance, evaluated for a range of heavy neutrino masses with simulation, is about 0.20 up to 360 MeV/$c^2$ and drops to zero for larger masses.
The statistical analysis is performed by
applying the Rolke-Lopez method~\cite{rolke} to find the 90\% confidence intervals on the number of reconstructed $K^+\to\mu^+\nu_h$ events
for the case of a Poisson process in the presence of Gaussian backgrounds. Inputs to the computation in each mass window are the number of data events observed, and
the estimate of the total number of background events with its uncertainty.
The squared uncertainties on the numbers of expected events in each mass hypothesis are shown in Figure~\ref{fig:syst},
where the various contributions can be seen.

\begin{figure}[htb]
\begin{minipage}[t]{18pc}
\includegraphics[width=18pc]{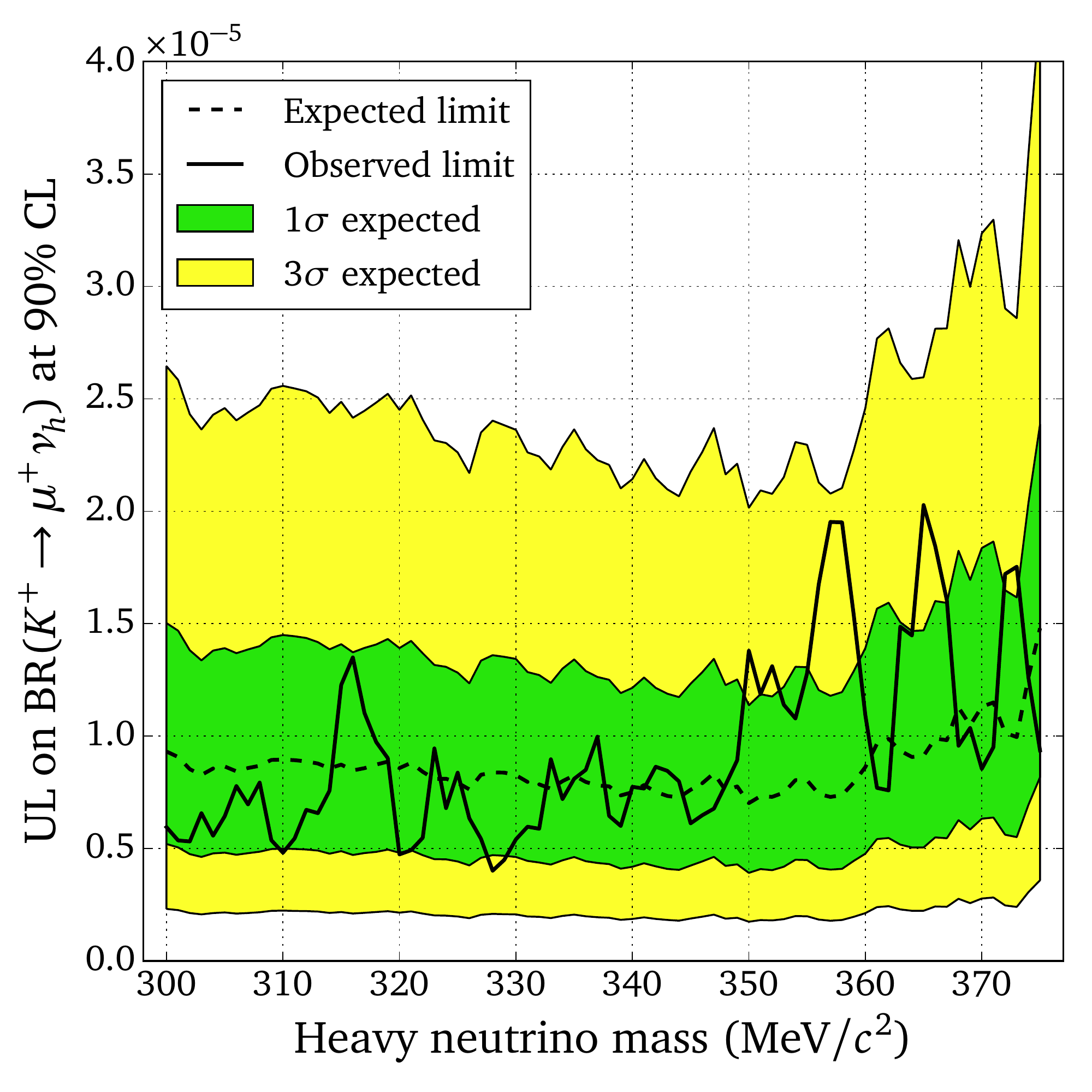}
\caption{\label{fig:branc} Expected and observed upper limits (at 90\% CL)  on the branching ratio in $10^{-5}$ units of the $K^+ \rightarrow \mu^+ \nu_{h}$ decay at each assumed $\nu_h$ mass.}
\end{minipage}\hspace{2pc}
\begin{minipage}[t]{18pc}
\includegraphics[width=18pc]{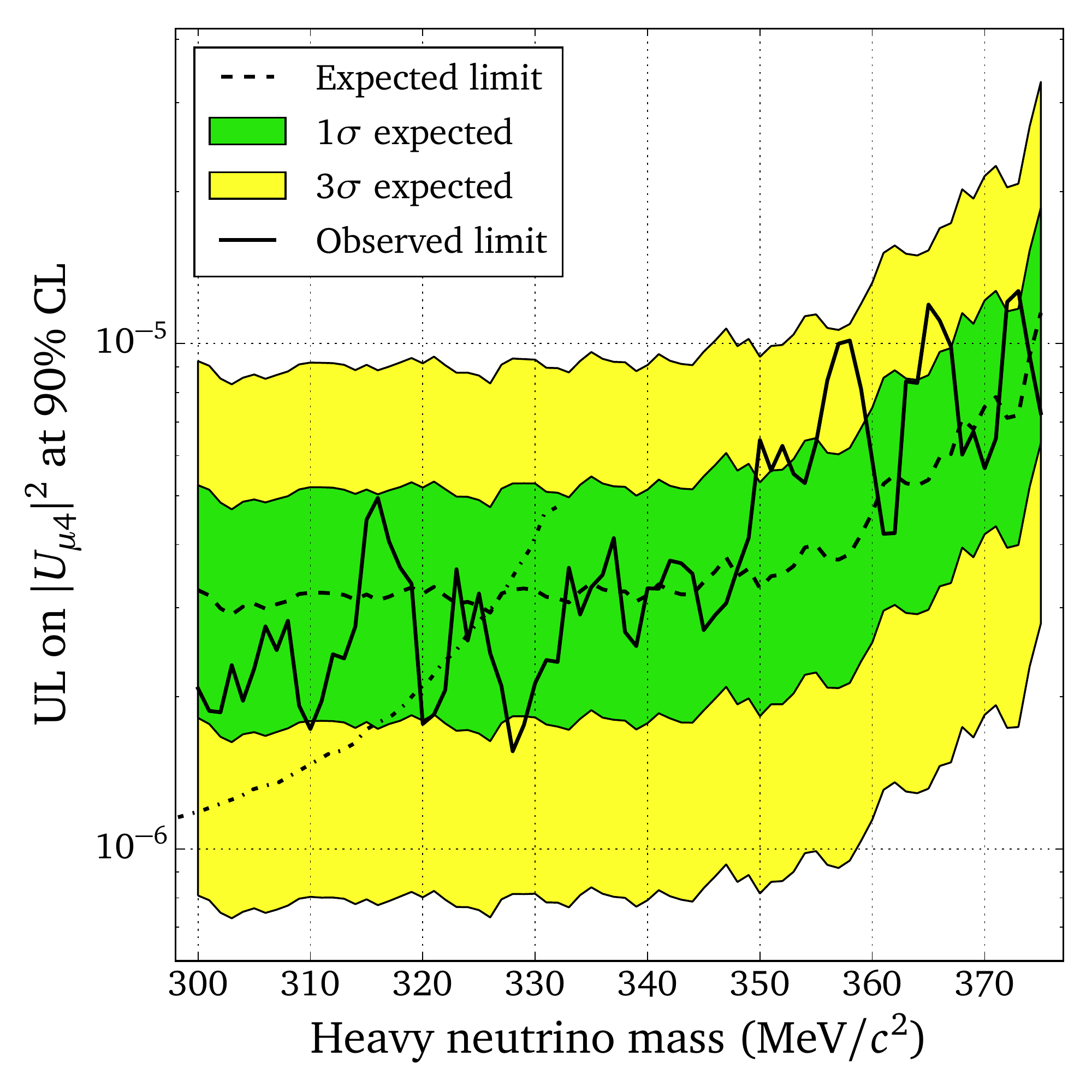}
\caption{\label{fig:u2} Expected and observed upper limits (at 90\% CL) on the matrix element squared $|U_{\mu 4} |^2$ at each assumed $\nu_h$ mass. The existing limit from KEK E089~\cite{hay} is also shown (dotted line). Below 300 MeV/$c^2$ there is a limit of $\mathcal{O}(10^{-8})$ from BNL E949~\cite{arta}, not shown. }
\end{minipage} 
\end{figure}

No signal is observed, the maximum local significance being 2.67 standard deviations at 357~MeV/$c^2$. 
The upper limits (UL) at 90\% CL on the numbers of reconstructed $K^+\to\mu^+\nu_h$ events is indicated as $n_{UL}$.
The expected upper limits are calculated assuming that the number of events observed is equal to the number of events expected, i.e. the number of background events. 
These upper limits are converted to upper limits on the branching ratio $\mathcal{B} (K^+ \rightarrow \mu^+ \nu_{h} )$ as shown in Figure~\ref{fig:branc}, using the relation $n_{UL} = \mathcal{B}_{UL} (K^+ \rightarrow \mu^+ \nu_{h}) A (m_h ) N_K$, where $A(m_h )$ is the signal acceptance, $ \mathcal{B}_{UL}$ is the upper limit on the branching ratio, and $N_K$ is given in Section~\ref{background}. 
The branching ratio is related to the neutrino mixing-matrix element squared $|U_{\mu 4}|^2$ by equation~(\ref{eq:formula}).
The obtained upper limits on  $|U_{\mu 4} |^2$ are shown in Figure~\ref{fig:u2}, together with the
limits from a previous peak search experiment~\cite{hay}.

\clearpage
\section{Conclusions}
A peak search has been performed in the missing mass spectrum observed in $K^+ \rightarrow \mu^+ N$ decays using part of the NA62 2007 dataset. Limits in the range $2 \times 10^{-6}$ to $10^{-5}$ have been set on the mixing matrix element squared between muon and heavy neutrino states for assumed neutrino masses in the range 300--375~MeV/$c^2$. 
The result extends the range of masses for which upper limits have been set on the value of $|U_{\mu 4}|^2$ by previous $\nu_h$ production experiments.
Thanks to the design and excellent performance of the current NA62 setup~\cite{na62_detector}, a substantial improvement in sensitivity is expected.

\section*{Acknowledgements}
We gratefully acknowledge the CERN SPS accelerator and beam-line staff for the excellent
performance of the beam and the technical staff of the participating institutes for their efforts
in the maintenance and operation of the detector, and data processing.
 



\begin{thebibliography}{99}
\bibitem{pal}
R.~N.~Mohapatra and P.~B.~Pal, World Sci. Lect. Notes Phys. {\bf 72} (2004) 1.
\bibitem{moh}
R.~N.~Mohapatra et al., Rept. Prog. Phys. {\bf 70} (2007) 1757. 
\bibitem{asaka}
T.~Asaka, S.~Blanchet, and M.~Shaposhnikov, Phys. Lett. B {\bf 63} (2005) 151.
\bibitem{shap}
T.~Asaka and M.~Shaposhnikov, Phys. Lett. B {\bf 620} (2005) 17.

\bibitem{shrock}
R.~Shrock, Phys. Lett. B {\bf 96} (1980) 159. 
\bibitem{hay}
R.~S.~Hayano et al. [KEK E089 Collaboration], Phys. Rev. Lett. {\bf 49} (1982) 1305.
\bibitem{arta}
A.~V.~Artamonov et al. [BNL E949 Collaboration], Phys. Rev. D {\bf 91} (2015)  052001. 
\bibitem{gorb}
D.~Gorbunov and M.~Shaposhnikov, JHEP  {\bf 0710} (2007)  015, Erratum JHEP {\bf 1311} (2013) 101.

\bibitem{detector}
J.~R.~Batley et al. [NA48/2 Collaboration], Eur. Phys. J. C {\bf 52} (2007) 875.
\bibitem{beam}
V.~Fanti et al. [NA48 Collaboration], Nucl. Instrum. Meth. A {\bf 574} (2007) 433.  

\bibitem{rk}
C.~Lazzeroni et al. [NA62 Collaboration], Phys. Lett. B {\bf 719} (2013) 326.  


\bibitem{rk1}
C.~Lazzeroni et al. [NA62 Collaboration], Phys. Lett. B {\bf 698} (2011) 105.
\bibitem{rolke}
W.~A.~Rolke and A.~M.~Lopez, Nucl. Instrum. Meth. A {\bf 458} (2001) 745.

\bibitem{na62_detector}
E.~Cortina~Gil et al. [NA62 Collaboration], 2017 JINST 12 P05025.

\end{thebibliography}
\end{document}